\documentclass[11pt]{article}
\usepackage{amsmath}
\usepackage{graphicx,psfrag,epsf}
\usepackage{enumerate}
\usepackage{natbib}
\usepackage{url}
\usepackage{ amssymb, amsthm, bm}
\usepackage{bmpsize}
\usepackage{psfrag, epsf}
\usepackage{enumerate}
\usepackage{lineno}
\modulolinenumbers[5]
\usepackage{booktabs}
\usepackage[utf8]{inputenc}
\usepackage{graphicx}
\usepackage{natbib}
\usepackage[breaklinks]{hyperref}
\usepackage{breakurl}
\usepackage{flafter}
\usepackage{float}
\usepackage{adjustbox}
\usepackage{tgbonum}

\newcommand{\blind}{0}

\usepackage{multicol}
\usepackage{adjustbox}
\usepackage{cclicenses}
\usepackage{makecell}
\usepackage{lscape,array}
\newcolumntype{C}[1]{>{\centering\arraybackslash}p{#1}} 
\usepackage[thin, , thinc]{esdiff}
\usepackage{subcaption}
\usepackage{placeins}
\usepackage{xcolor}
\usepackage{istgame}
\usepackage{wrapfig}
\usepackage{bmpsize}
\usepackage{flafter}
\usepackage{color, hyperref, xcolor}
\hypersetup{colorlinks=true, urlcolor=blue, citecolor=purple}
\usepackage{cleveref}
\numberwithin{equation}{section}

\newtheoremstyle{general}
{3mm} 
{3mm} 
{} 
{} 
{\bfseries} 
{.} 
{.5em} 
{} 

\theoremstyle{general}

\begin{document}
\bibliographystyle{chicago}

\def\spacingset#1{\renewcommand{\baselinestretch}%
{#1}\small\normalsize} \spacingset{1}


\if0\blind
{
  \title{Analysis and Estimation of Consumer Expenditure assuming uniform prices across FSUs:  A step towards cost effectiveness}
\author{Rajdeep Brahma, Anagh Chattopadhyay,\\ Diganta Mukherjee \& Tathagata Sadhukhan\thanks{Corresponding author, email: tathagatasadhukhan@gmail.com}    \vspace{.1cm}\\
  Indian Statistical Institute, 203 B.T. Road, Kolkata 700108, India.}
  \maketitle
} \fi

\if1\blind
{
  \bigskip
  \bigskip
  \bigskip
  \begin{center}
    {\LARGE\bf Title}
\end{center}
  \medskip
} \fi

\bigskip
\begin{abstract}
Analysis and estimation of consumer expenditure and budget shares are important for understanding quantitatively the expenditure based behaviour of the people of a country or region. The costs attached with performing consumer expenditure and budget shares survey are significant. These surveys are quite time consuming and even a slight increase in sample size could lead to a significant increase in the cost of the survey under the current structure. This high cost for the survey also reduces its flexibility. Hence it is of paramount importance to be able to provide a statistically sound and relatively cheaper facilitation method of survey to be able to understand the consumption expenditure distribution of a country or a region without much loss of inferential power. In the context of the Indian National Sample Survey, in this paper we perform analysis and estimation of consumer expenditure assuming uniform prices across First Stage Units (FSUs) of a sampling design, and check its feasibility in estimating the consumer expenditure distribution of the country. We also compare it with the existing methodology and infer that there is no significant loss of information in the estimated consumer expenditure distribution and other inferences like the Lorenz curve and the Gini Index, when uniform distribution of prices within any FSU is assumed.
\end{abstract}

\noindent%
\vfill

\newpage
\spacingset{1.8} 
\section{Introduction} \label{sec:introduction}
Analysis and estimation of consumer expenditure and budget shares are an important process which helps us understand quantitatively the expenditure based behaviour of the people of a country or region. Analysis of consumption expenditure is important for understanding short term as well as long term economic phenomena because consumption expenditure to a huge degree can demonstrate fluctuations in various important economic factors which in turn helps us perceive it as an expression of economic stability and health. Even a small decrease in consumption expenditure can damage the economy and hence it is quite important to be able to understand and precisely quantify as well as analyse the consumption expenditure distribution of a country or a region of interest. \cite{krishnaswamy2012pattern} provide excellent examples of the utility of expenditure surveys. The paper comments on the various insights which can be drawn from the consumer expenditure surveys, such as the growth of GDP and other relevant economic factors.       

\cite{sen2000estimates}, \cite{ghosh2011consumer} and \cite{sen2001consumer} are important papers which touch upon the subjects of distribution of the consumer expenditure of the people of India. These papers emphasize the need of statistical analysis of consumer expenditure behaviour and study the various facets of the consumer expenditure distribution and their implications towards the economy and the society in general. These papers also touch upon the methodology corresponding to the collection of data through the surveys conducted by the Indian National Sample Survey Organization (NSSO) and also gives us an idea about the shortcomings and the cost of such surveys. \cite{sen2000estimates} also mentions the costs attached while performing consumer expenditure and budget shares survey and its implications towards statistical reliability,  because with the current setup of collecting a large amount of data from every household,  achieving statistically sound survey data is often quite time consuming and an increase in sample size would lead to a huge increase in the cost of the survey under the current design. Hence it is of paramount importance to be able to provide a statistically sound and relatively cheaper method for facilitation of survey, to be able to understand the consumption expenditure distribution of a country or a region without much loss of inferential strength. \cite{seshadri_2019} states that in 2017, the CES report was withheld citing bad data quality. The Ministry said “Concerns were also raised about the ability/sensitivity of the survey instrument to capture consumption of social services by households especially on health and education. The matter was thus referred to a committee of experts which noted the discrepancies and came out with several recommendations including a refinement in the survey methodology.” Hence the need of a robust and cheap method of survey with a sound statistical basis is vital.        

Understanding and modelling of consumer expenditure as well as the corresponding budget share has been an important part of the economic literature. Many models have been proposed to analyse and understand the budget share corresponding to various commodities for a given family or households.   
\cite{deaton_muellbauer_1980} provides an excellent review of the various tools which have been used throughout the years for consumer expenditure and budget share estimation and analysis. The various models range from simple classical models such as the linear expenditure system or the Rotterdam system to more complex models such as the Almost Ideal Demand System (AIDS model). Applications of the AIDS model is quite popular for studying consumption behaviour and consumption modelling and has been widely used as in \cite{syriopoulos1993econometric},  \cite{fan1995household},  \cite{yang1994japanese} and \cite{ahmed1994demand} which presents many interesting applications of the model and highlights many of its utilities.       

We have also introduced  measurement error correction in our model structure under a cross validation setup to be able to generalise our observations by incorporating errors in measurement in both the prices observed of the various items considered in the surveys and total expenditure, using tools described in \cite{Nab2021-ts} and \cite{nab2021sensitivity}.        \cite{berger2000bounding},  \cite{klepper1984consistent} and \cite{hyslop2001bias} provide the conceptual framework and also outlines the necessity for inclusion of classical error measurement and its analysis with respect to our model and creation of the corresponding statistical framework. \cite{griliches1986errors} also inspires a lot of the analysis performed here towards creation of a generalised model with more flexibility and interpretability.       

In this paper we perform analysis and estimation of consumer expenditure assuming uniform prices across FSU's. We can easily understand that if we assume uniform prices across FSU's then the amount of data that would be required in the survey would be much less than in the case of every household in the FSU being asked about the prices. Hence we try to understand the statistical reliability of replacing all the individual values of prices in a given FSU by one of the prices from the set of households, considering it as a representative of the prices for the entire FSU. In this paper a modified version of the AIDS model,  considering measurement errors in data collected corresponding to prices as well as consumption expenditure,  has been used to understand the dissimilarities/similarities in the obtained budget shares and total expenditures in both the usual approach of complete data collection and the approach with assuming uniform prices across FSUs. A review of the relevant literature is made in the next section. The description of data, methodology and results obtained under alternative assumptions are discussed over several subsections of section 3. In particular, we analyse the robustness of our results in the presence of measurement error in a very general setting.

Apart from the budget share, two other very important metrics in the context of the CEs are measurement of inequality and demand elasticity. The Lorenz curve and the Gini Index are two most well known measures of inequality. Section 4 looks at the related issue of the estimation of the Lorenz curve and the Gini Index in the usual as well as our proposed approach. Similar analysis for demand elasticity is carried out in Section 5. Finally section 6 concludes the paper.

\section{Literature Review}
Consumer Expenditure models (CEMs) are widely studied due to the interpretation it provides to the household expenditure as a function of various budget shares and prices of commodities. Various economic models for consumer analysis have been proposed and implemented over the years. One of the first empirical model  was the Working-Leser model. The original form of the Working-Leser model was discussed by \cite{working1943statistical} and \cite{leser1963forms},  with recent implementation in \cite{karunakaran1996use},  \cite{brosig2000model},  \cite{chern2002analysis} and \cite{yeong2008household}. \cite{intriligator1996econometric} and \cite{deaton_muellbauer_1980} also discuss these functional forms as well as describe modifications and implementations on real life data. \cite{amemiya1985instrumental} and \cite{maddala1983methods} study the Tobit and Heckman's two-step estimator \cite{heckman2003simple},  which is another important estimator which improves on the Working-Leser model structure.        \cite{mcdonald1980uses} also deals with this model and improves upon the understanding of the model structure.       

\cite{deaton_muellbauer_1980} created an efficient and flexible demand system AIDS,  which stands for 'Almost Ideal Demand System'. It is not just extremely popular but also extremely versatile. The notion of a flexible demand system comes in handy when evaluating a demand system with a lot of desired characteristics. The AIDS model, as \cite{moschini1998semiflexible} and \cite{henningsen2017demand} point out,  meets the aggregation limitation,  and provides facility to test and impose the homogeneity and symmetry with simple parametric requirements. Furthermore,  the non-linear Engel curves of the AIDS model suggest that when income rises,  the percentage of income devoted to a given commodity decreases,  as does the income elasticity of that product when it is less than one. It also satisfies the axioms of choice,  is an arbitrary first-order approximation to any demand system. The functional form of AIDS is still widely used in recent publications (see, e.g. \cite{brannlund2007increased}; \cite{chambwera2007fuel}; \cite{farrell2007accuracy}; and also \cite{ahmed1994demand} among many others)

\cite{henningsen2017demand} is an important contribution towards implementation of the AIDS models as it itself states -"The aim of this paper is to describe the tools for demand analysis with the AIDS that are
available in the R (R Development Core Team 2008) package micEconAids (version 0.5)."  \cite{henningsen2017demand} implements both the AIDS as well as the LA-AIDS models in R while highlighting modifications.        This paper uses this package extensively for its usefulness and efficient functionalities.        

Measurement error was also included in this study to be able to generalise the model structure and widen the scope of the analysis. \cite{fuller2009measurement} is an excellent primer in the field of error measurement analysis. \cite{griliches1986errors} provides the R framework towards incorporating measurement errors in the consumer expenditure model structure.  \cite{nab2021sensitivity} is one of the most recent works in this domain. Some important works in this domain are \cite{angrist1991does},   \cite{aigner1973regression}, \cite{card1996effect}, \cite{ashenfelter1994estimates} and \cite{berger2000bounding}. Some other notable publications include on various applications of measurement error based methodolgies are \cite{bound1994evidence},  \cite{kane1999estimating}.  \cite{klepper1984consistent} and \cite{hyslop2001bias} are some important works in this field and provide excellent insights toward applications of measurement error analysis.        

\cite{wahba1980some} provides motivation towards Variational objective analysis using splines and cross validation. \cite{Pischke_measurement2007} is an excellent source, and provides a comprehensive primer towards understanding Measurement errors. We used the package \cite{nab2021mecor} for the utilisation of various measurement error based methodologies into our model structure.       

As defined by the U.S. Bureau of Labor Statistics \cite{Bureau_statUS}, "A consumer expenditure survey is a specialized study in which the emphasis is on data related to family expenditures for goods and services used in day-to-day living. In addition to data on family expenditures, the Consumer Expenditure Survey (CES) collects information on the amount and sources of family income, changes in assets and liabilities, and demographic and economic characteristics of family members".  \cite{Bureau_statUS} also states the importance of CES by stating "Data also are used by economic policymakers interested in the effects of policy changes on levels of living among diverse socioeconomic groups, and econometricians find the data useful in constructing economic models."

The utility of CEMs in understanding economic progress as well as growth quantitatively is unquestionable. Many papers, icluding recent ones, have analysed and inferred upon the state of various parameters which influence the economy, using CEMs as a primary tool in one form or the other, such as \cite{castro2012accommodating}, \cite{sundaram2021determinants}, \cite{meng2021determinants}, \cite{dobrowolska2021beef}, \cite{sakai2021analyzing}, \cite{kovacs2021consumer}, \cite{cacholacarbon}, \cite{dong2021dynamic} and many others.

\cite{chandra2021district},  \cite{sen2000estimates},  \cite{ghosh2011consumer},  \cite{sen2001consumer} highlight applications and usages of CEs in the Indian scenario, that is towards understanding the growth of the economy along with various other factors which directly affect our economy. It also shows us the changes in the economic structure across various classes of the society over a period of time, and implementations of CEMs helps us quantify the changes which occur in not just in terms of the expediture of the household, but also how the household chooses to use its income on consumption based expenditures, which in turn helps us to gain a deeper insight into the state of the Indian Economy.       

\section{The Demand System}
\label{sec:methods}

\subsection{Data}
The data was collected from \cite{NSS_2016}-" Household Consumer Expenditure: NSS 68th Round, Schedule 1.0, July 2011 - June 2012 (type 1 )."
As \cite{NSS_2016} states-"The programme of quinquennial surveys on consumer expenditure and employment \& unemployment, adopted by the National Sample Survey Organisation (NSSO) since 1972-73, provides a time series of household consumer expenditure data. Consumer expenditure surveys conducted in NSS rounds besides the 'quinquennial rounds' - starting from the 42nd round (July 1986 - June 1987) - also provide data on the subject for the period between successive quinquennial rounds, using a much smaller sample. The sixth instance of the quinquennial series was held during the 55th round (July 1999-June 2000). The seventh was conducted in the 61st round during July 2004 - June 2005 and 66th round of NSS (2009-10) which was the eighth quinquennial survey in the series on 'household consumer expenditure' and 'employment and unemployment'."

The data is composed of many blocks, which include
\begin{itemize}
\item Block 1 : Identification of sample household.       

\item Block 2 : Particulars of field operations.       

\item Block 3 : Household characteristics.       

\item Block 4 : Demographic and other particulars of household members.       

\item {\bf Block 5.1 : Consumption of cereals, pulses, milk and milk products, sugar and salt.}       

\item {\bf Block 5.2 : Consumption of edible oil, egg, fish and meat, vegetables, fruits, spices, beverages and processed food and pan, tobacco and intoxicants.}       
\item Block 6 : Consumption of energy (fuel, light \& household appliances).

\item Block 7 : Consumption of clothing, bedding, etc.       

\item Block 8 : Consumption of footwear.       

\item Block 9 : Expenditure on education and medical (institutional) goods and services.       

\item Block 10 : Expenditure on miscellaneous goods and services including medical (non-institutional), rents and taxes.       

\item Block 11 : Expenditure for purchase and construction (including repair and maintenance) of durable goods for domestic use.       
\end{itemize} and so on. We basically focus on block 5 (5.1 \& 5.2) in this study. There are 23 items that are considered in our analysis (see Table \ref{table:item name}). There were two data components for each item, one for urban households and one for rural households. The data also consist of The household IDs (denoted by HHID), the state from which the family hails (denoted by STATE), the net quantity consumed by the household (denoted by NHQ), the net household value (denoted by NHV), the monthly per capita expenditure (denoted by MPCE), and the multiplier (denoted by MULT).       
We combine these data into a single data structure that contains all of the households' ids, states, sizes, and shares of various items used by the household, as well as their prices.        

\begin{table}[!htp]
\begin{center}
\begin{tabular}{||c c||} 
 \hline
 item name & item number \\ [1ex] 
 \hline\hline
 Rice-PDS & i101 \\
 \hline
 Rice-other sources & i102  \\ 
 \hline
 Wheat/atta - other sources & i108  \\
 \hline
 suji, rawa & i111 \\
 \hline
 jowar, products & i115 \\
 \hline
 arhar (tur) & i140 \\
 \hline
 moong & i143 \\
 \hline
 masur & i144 \\
 \hline
 urd & i145 \\
 \hline
 besan & i151 \\
 \hline
 milk : liquid & i160 \\
 \hline
 ghee & i164 \\
 \hline
 salt & i170 \\
 \hline
 sugar - other sources & i172 \\
 \hline
 mustard oil & i181 \\
 \hline
 refined oil [sunflower,soyabean,saffola, etc] & i184 \\
 \hline
 goat meat /mutton & i192 \\
 \hline
 chicken & i195 \\
 \hline
 potato & i200 \\
 \hline
 onion & i201 \\
 \hline
 tomato & i202 \\
 \hline
 banana & i220 \\
 \hline
 tea : cups & i271 \\ [1ex] 
 \hline
\end{tabular}
\end{center}
\caption{ Items and their corresponding identification numbers}
\label{table:item name}
\end{table}

\subsection{Fitting the model}
The objectives of this paper are as follows:
\begin{enumerate}
    \item Fitting AIDS model to our data and estimating expected shares from this model.
    \begin{equation}
 \label{eq:AIDS model}
w_{il}=\alpha_i+\sum_{j} \gamma_{ij} *log(p_j)+\beta_i*log(x_{l}/P)+\epsilon_{il}
\end{equation}
for the $l^{th}$ household, under the restriction $\sum_{i}^{} \alpha_i=1$, $\sum_{i}^{} \beta_i=0$, $ \sum_{i}^{} \gamma_{ij}=0$, $ \sum_{j}^{} \gamma_{ij}=0$ and $\gamma_{ij}=\gamma_{ji}$ for all i and j. 
    \item Creating an alternate data set from the original one by making the price of an item constant in each FSU and checking the distributional similarity of shares with the one estimated above.
    \item Checking the same distributional similarity as explained in 2, taking into account state effects. Hence this provides a generalisation to the model setup already used for structuring the consumer expenditure data.    
    \item Checking the same distributional similarity as explained in 2, taking the total expenditure and prices as random effects. This provides a further improvement to our model setup, and helps us further obtain a generalised data structure.       
\end{enumerate}

We use R package \textbf{"micEconAids"} to fit the above model. One can refer to papers such as \cite{henningsen2011miceconaids} and \cite{henningsen2017demand} for better understanding of the package and its specifications.       

\subsection{Distributional similarity of shares under the assumption of equality of prices in each FSU}
Households with same HHIDs except the last 2 digits falls under the same FSU.   First we fit our AIDS model to the original dataset. Now micEconAids (\cite{henningsen2017demand}) package drops a row of the data matrix even if one element of it is NA. We have chosen a pool of 6 items with the least amount of NA entries  in them among all the 23 items (2.4\%, 5.3\%, 9.7\%, 9.9\%, 12.8\% and 14\% of the entries corresponding to these items are NA respectively) using which we can estimate their shares most accurately by this package. The items turned out to be \emph{i202, i160, i108, i143, i220} and \emph{i200}. We find their expected shares $(w_i^{1})$. Next we create a new dataset where all the item prices in each FSU is constant. We refit our model for this new dataset using only the above mentioned items and evaluate the expected shares$(w_i^{2})$.     We check if $w_i^{1}$ has the same distribution as $w_i^{2}$. 

We do a 2-sample both sided Kolmogorov-Smirnov (KS) test on the individual components of $w_i's$. There are 6 consumption items. For \emph{i160}, \emph{i143} and \emph{i200} the hypothesis was accepted and in all other case it was rejected at 5 \% significance level. The results obtained for the KS test are shown in table \ref{table:7} (columns 3 \& 4).       
\begin{table}[H]
\begin{center}
\begin{tabular}{||c c | c c|cc||} 
  \hline
&& \multicolumn{2}{|c|}{w.o. state effect} & \multicolumn{2}{|c||}{with state effect} \\
\hline
item & D & p-value & Accept/Reject & p-value & Accept/Reject \\ [2ex] 
\hline \hline
i108  & 0.043197  &  $1.099\times10^{-7}$ &  Reject & 0.4944  & Accept \\
\hline
i143  & 0.012725  & 0.4628  & Accept & 0.3086  & Accept   \\
\hline
i160  & 0.012167  & 0.5221  & Accept & 0.7811  & Accept   \\
\hline
i200  & 0.012948  & 0.4405  & Accept   & 0.4229  & Accept \\
\hline
i202  & 0.028575  & 0.001331  & Reject & 0.0007453  & Reject  \\
\hline
i220  & 0.05101  & $1.05\times10^{-7}$  & Reject & 0.4821  & Accept   \\
\hline
\end{tabular}
\end{center}
\caption{K-S test with and without state effect}
\label{table:7}
\end{table}

Next we did the same hypothesis testing on the entire share vectors rather than component wise by Cramer-test for multivariate two-sample problem. It was also rejected and so we can claim that making the prices same in all the houses of the FSU do not keep the distribution of expected share the same.        

\subsection{Inclusion of state effects}
Inclusion of the spatial properties of consumer expenditure is an important part of the process of generalising the model for households in the data set. The main reason for the inclusion of spatial properties into the model is the fact that the consumer expenditure for various items vary from one region to another owing to people from different regions having different preferences with regards to their consumption and also due to the different practices corresponding to food and other essentials in different parts of the country. To imbibe the spatial structure into the model we consider the state effects as a factor in our model which helps us to differentiate between the consumption patterns for different states. We follow the philosophy described earlier pertaining to the use of non parametric tests for the equality of the share distribution under the generalized model structure as follows       
\begin{equation}
 \label{eq:AIDS model generalised(1)}
w_{il}=\alpha_i+\sum_{j}^{} \gamma_{ij} *log(p_j)+\beta_i*log(x_{l}/P)+\S_{kl}+\epsilon'_{ikl}
\end{equation}
where $\S_{kl}$ is the effect of the state corresponding to the k$^{th}$ state, and l$^{th}$ household.       
 To find the coefficients $\S_{k}$,  we partially regress the error terms $\epsilon_{ik}$ (the residuals in eq. (\ref{eq:AIDS model}) with an extra index for state) w.r.t. the state effects as a factor, and we substitute the coefficients obtained from the partial regression on $\epsilon_{ik}$, 
$\epsilon_{ikl}=\S_{kl}+\epsilon'_{ikl}$ 
in eq.(\ref{eq:AIDS model}), to obtain the AIDS model with state effect giving us new predicted shares $w_i'$. We utilise this method to assure the fact that the restrictions are upheld in the given scenario.

The results obtained for the KS test for the same setup as before, that is using items \emph{i202, i160, i108, i143, i220} and \emph{i200}, are shown in table \ref{table:7} (columns 5 \& 6). We now obtain extremely motivating results as 5 out of 6 items are accepted as having the same budget share distribution according to the two-sided tests

It is understood that State heterogeneity contributed a lot towards the rejection of the budget share distributional equality hypothesis in the earlier analysis.

\subsection{Measurement Error only in Total Expenditure:}

Consumption and expenditure data is often subject to measurement error, as already discussed in section 2. To study the effect of this feature on the conclusion of our analysis, in this section we first include measurement error in the total expenditure variable only. Suppose $x_i^{*}$ is the original expenditure of the i-th household but the reported expenditure is $x_i$. In terms of a linear model we express it as 
\begin{equation}
log(\frac{x_i^{*}}{P})=\theta_0+\theta_1 log(\frac{x_i}{P})+\epsilon_i 
\end{equation}
where $\Theta=(\theta_0, \theta_1)$ and i refers to the household. We now incorporate this model in our original AIDS model and calculate the parameters for that $\Theta$ vector which has the least error.       

\subsubsection{Estimation of parameters}
\label{subsec:est of para}
Suppose our original model is 
\begin{equation}
y|(X, Z)= \beta_0+ \beta_X X+\beta_Z Z + \epsilon 
\end{equation}

But we have data on responses $Y$ and error-prone predictors $\Tilde{X}$ where $\Tilde{X}=\theta_0+\theta_1 X + U$ where $U$ is the measurement error matrix. Assume $E(U)=0$ and $Cov(U, X)=0.$ We assume non-differential covariate measurement error. The measurement error is called ‘classical’ or
‘random’ if $\theta_0 = 0$ and $\theta_1 = 1$.  In this paper, we use the
term random measurement error to refer to this type of measurement error.
Now let 
\begin{equation}
\label{eqn:measure1}
y|(\Tilde{X}, Z)=\Tilde{\beta_0}+  \Tilde{\beta_X} \Tilde{X}+\Tilde{\beta_Z} Z + \Tilde{\epsilon}
\end{equation} The least square estimators of \eqref{eqn:measure1},   $\hat{\Tilde{\beta}}=(\hat{\Tilde{\beta}}_X,\hat{\Tilde{\beta}}_0,\hat{\Tilde{\beta}}_Z)$ are biased for $\beta$ but are consistent and unbiased estimators for $\beta\Lambda$ where $\Lambda$ is the calibration model matrix (see \cite{buonaccorsi_2010}, \cite{nab2021mecor}).

\cite{nab2021mecor} derives the form for the calibration model matrix as below.
\begin{equation*}
\label{lamb}
    \Lambda=\begin{bmatrix}
\lambda_{\Tilde{X}} & \lambda_0 & \lambda_Z\\
0 & \multicolumn{2}{c}{I}
\end{bmatrix}
\end{equation*}

 where $I$ is a $(k+1)*(k+1)$ identity matrix, 0 is a $(1+k)*1$ null matrix and $\lambda_Z$ is a $1*k$ matrix where $X$ is a k length vector and $Z$ is a $k$ length vector.  It follows that $\hat{\Tilde{\beta}}\Lambda^{-1}$ are consistent and unbiased estimators for $\beta.$ This understanding helps us to derive the $\Lambda$ matrix in terms of $\theta_0$ and $\theta_1$ as follows.

From \eqref{eq:AIDS model} we have 
\begin{equation}
\begin{split}
     E(w_{il})=\Tilde{\alpha_i}+\sum_{j}^{} \Tilde{\gamma_{ij}} *log(p_j)+E(\Tilde{\beta_i}*log(\Tilde{x_{l}}/P))\\
    =\Tilde{\alpha_i}+\sum_{j}^{} \Tilde{\gamma_{ij}} *log(p_j)+\Tilde{\beta_i}*(\theta_0+\theta_1 log(\frac{x_i}{P}))
    \\={\alpha_i}+\sum_{j}^{} \gamma_{ij} *log(p_j)+{\beta_i}*log({x_{l}}/P)
\end{split}
\end{equation}

From here we can claim that $\Tilde{\alpha_i}+\theta_0 \Tilde{\beta_i}=\alpha_i,\Tilde{\gamma_{ij}}=\gamma_{ij}$ and $\Tilde{\beta_i}\theta_1=\beta_1$
\\Accordingly the calibration model matrix becomes $
\Lambda=\begin{bmatrix}
\frac{1}{\theta_1} & -\frac{\theta_0}{\theta_1} & 0\\
0 & \multicolumn{2}{c}{I}
\end{bmatrix}$

An important special case of the calibration model matrix is where we have only one response and one error prone predictor. In those situations we have $\Tilde{\beta_X}=\lambda_{\Tilde{X}}\beta_X$. Then $\lambda_{\Tilde{X}}$ is called the attenuation factor or regression dilution factor.  To be able to expand this idea and for deriving the optimal values for $\theta_0$ as well as $\theta_1$, we apply a cross-validation based approach. Cross-validation is one of the most widely used data resampling methods to estimate the true prediction error of models and to tune model parameters. \cite{stone1978cross} and \cite{refaeilzadeh2009cross} are some excellent reviews of various cross-validation techniques and approaches.

We utilise 10 fold cross-validation for deriving the optimum values for $\theta_0$ and $\theta_1$. We divide the data into 10 equal parts. From $\frac{9}{10}$th of the data we estimate the values of parameters $(\beta_0,\beta_X,\beta_Z)$ and then on the remaining $\frac{1}{10}$th of data we apply the trained model, obtain the estimated shares in each household and then we compute the Cross validation (CV) error as a measure of error or divergence between the estimated and the true shares. This Cross validation (CV) error is very useful for deriving or quantifying a loss function which will help us derive the values for $\theta_0$ as well as $\theta_1$ which provide us the best model parameters. The loss functions that we will be using in this context are $L_1$ and $L_2$, to implement alternative sensitivity to outliers.

\subsubsection{Results}
Interestingly, we find  that the value of $\theta_0$ didn't affect the CV error much so only the change of CV error w.r.t. the value of $\theta_1$ is reported here while keeping $\theta_0$ at 0. The set of items with minimum number of NA entries among them are i170, i200, i201 and i271. Results were obtained for these items, as shown in table \ref{table:0}. Best results were obtained for $\theta_1 = 1.02$ which is quite close to $1$ and since $\theta_0=0,$ so we can conclude that the measurement error in total expenditure is quite close to the scenario of "classical measurement error"(\ref{subsec:est of para}) i.e., measurement error only arises from random noise and not from any biased nature in observed total expenditures. So, the inclusion of measurement error for correcting the bias in the reported total expenditure hardly shows any improvement in the analysis.

\begin{table}[!htp]
    \centering
    \begin{tabular}{||c| c |c||}
    \hline
        $\theta_1$ & CV error (L1) & CV error (L2)\\
        \hline \hline
  0.5 & 873.8071 & 587.8226 \\
  0.6 & 765.0441 & 490.4459\\
  0.7 & 582.5900 & 354.2595\\
  0.8 & 507.0505 & 305.7418\\
  0.9 & 4 25.4642 & 249.4999\\
  0.92 & 396.9889 & 222.655   \\
  0.94 & 388.7303 & 211.9089\\
  0.96 & 390.3062 & 207.9725\\
 0.98 & 385.0477 & 202.5024\\
    1 & 382.3179 & 180.2719\\
    \hline
    1.02 & 381.3437 & 178.9530\\
 \hline
 1.04 & 382.5989 & 181.904\\
 1.06 & 383.2434 & 186.2684\\
 1.08 & 383.8451 & 186.6432\\
  1.1 & 386.8990 & 196.8482\\
  1.2 & 389.7244 & 198.0584\\
  1.3 & 425.8812 & 227.3775\\
  1.4 & 460.3024 & 265.6682\\
  1.5 & 574.4676 & 352.5212\\
  \hline
    \end{tabular}
\caption{CV for measurement error analysed only for total expenditure}
\label{table:0}
\end{table}

\subsection{Measurement Error in both Prices and Expenditure:}

In this section, we include measurement error in total expenditure as well as prices.  In terms of a linear model we express it as 
\begin{equation}
\begin{split}
log(\frac{x_i^{*}}{P})=\theta_0+\theta_1
log(\frac{x_i}{P})+\epsilon_i
\\
log(p_{ik}^{*})=\theta_2+\theta_3 log(p_{ik})+\eta_{ik}
\end{split}
\end{equation}
where $\Theta=(\theta_0, \theta_1, \theta_2, \theta_3)$, i and k refers to the household and item respectively. We next incorporate this model in our original AIDS model and obtain the parameters for that $\Theta$ vector which has the least error.

\subsubsection{Estimation of parameters:}
Similar to the previous section, From (\ref{eq:AIDS model}) we have 
\begin{equation}
\begin{split}
     E(w_{il})=\Tilde{\alpha_i}+\sum_{j}^{} E(\Tilde{\gamma_{ij}} *log(\Tilde{p_j}))+E(\Tilde{\beta_i}*log(\Tilde{x_{l}}/P))\\
    =\Tilde{\alpha_i}+\sum_{j}^{} \Tilde{\gamma_{ij}} *(\theta_2+\theta_3 log(p_{ik}))+\Tilde{\beta_i}*(\theta_0+\theta_1 log(\frac{x_i}{P}))
    \\={\alpha_i}+\sum_{j}^{} \gamma_{ij} *log(p_j)+{\beta_i}*log({x_{l}}/P)
\end{split}
\end{equation}

Hence we claim that $\Tilde{\alpha_i}=\alpha_i-\frac{\theta_0}{\theta_1}\beta_i-\frac{\theta_2}{\theta_3}\sum_{j}\gamma_{ij},\Tilde{\gamma_{ij}}=\frac{\gamma_{ij}}{\theta_3}$
and $\Tilde{\beta_i}\theta_1=\beta_1.       $
\\Accordingly the calibration model matrix becomes 
$$\Lambda=\begin{bmatrix}
1 & 0 & 0\\
-\frac{\theta_0}{\theta_1} & \frac{1}{\theta_1} & 0\\
-\frac{\theta_2}{\theta_3}1 & 0 & \frac{1}{\theta_3}I_k
\end{bmatrix}$$

Now again following the Cross-Validation technique as explained before, we calculate the CV error as shown in tables \ref{table:3}, \ref{table:4}, \ref{table:5} and \ref{table:6}. Since in previous section we had obtained the result that effect of measurement error on total expenditure is negligible, we fixed $\theta_1=1$.       

\begin{table}[!htp]
\begin{center}
\begin{tabular}{||c c c c||} 
 \hline
$\theta_1$ & $\theta_2$ & $\theta_3$ & CV Error \\ [2ex] 
 \hline\hline
1  & 0.045  & 0.975  & 390.7835   \\
\hline
1  & 0.03  & 0.975  & 388.8978   \\
\hline
1  & 0.015  & 0.975  & 386.4447   \\
\hline
1  & 0  & 0.975  & 387.6745   \\
\hline
1  & -0.015  & 0.975  & 387.7026   \\
\hline
1  & -0.03  & 0.975  & 387.9214   \\
\hline
1  & -0.045  & 0.975  & 388.3474  \\
\hline
1  & 0.045  & 0.9825  & 384.017   \\
\hline
1  & 0.03  & 0.9825  & 383.0928   \\
\hline
1  & 0.015  & 0.9825  & 385.5355   \\
\hline
1  & 0 & 0.9825  & 392.1737   \\
\hline
1  & -0.015  & 0.9825  & 385.2428   \\
\hline
1  & -0.03  & 0.9825  & 386.7386   \\
\hline
1  & -0.045  & 0.9825  & 384.7865   \\
\hline
1  & 0.045  & 1  & 382.3325   \\
\hline
1  & 0.03  & 1  & 383.0103   \\
\hline
1  & 0.015  & 1  & 383.008   \\
\hline

1  & 0  & 1  & 384.6481   \\
\hline
\end{tabular}
\begin{tabular}{||c c c c||} 
 \hline
$\theta_1$ & $\theta_2$ & $\theta_3$ & CV Error \\ [2ex] 
 \hline\hline
1  & -0.015  & 1  & 386.7948   \\
\hline
1  & -0.03  & 1  & 385.7552   \\
\hline
1  & -0.045  & 1  & 385.1635   \\
\hline

1  & 0.045  & 1.0125  & 385.2014   \\
\hline
1  & 0.03  & 1.0125  & 382.8658   \\
\hline
1  & 0.015  & 1.0125  & 383.7602   \\
\hline
1  & 0  & 1.0125  & 383.4262   \\
\hline
1  & -0.015  & 1.0125  & 383.9514   \\
\hline
1  & -0.03  & 1.0125  & 391.0214   \\
\hline
1  & -0.045  & 1.0125  & 382.3354   \\
\hline
1  & 0.045  & 1.025  & 383.0567   \\
\hline
1  & 0.03  & 1.025  & 383.3367   \\
\hline
1  & 0.015  & 1.025  & 389.2024   \\
\hline
1  & 0  & 1.025  & 383.2446   \\
\hline
1  & -0.015  & 1.025  & 384.9549   \\
\hline\hline
1  & -0.03  & 1.025  & 382.2439   \\
\hline\hline
1  & -0.045  & 1.025  & 386.6325  \\
\hline
 \hline
\end{tabular}
\end{center}
\renewcommand\thetable{4A}
\caption{CV L1 error (item set: i170, i200, i201, i271)}
\label{table:3}
\end{table}
\begin{table}[H]
\begin{center}
\begin{tabular}{||c c c c||} 
 \hline
$\theta_1$ & $\theta_2$ & $\theta_3$ & CV Error \\ [2ex] 
 \hline\hline
1  & 0.045  & 0.975  & 208.0036   \\
\hline
1  & 0.03  & 0.975  & 205.6403   \\
\hline
1  & 0.015  & 0.975  & 197.3612   \\
\hline
1  & 0  & 0.975  & 202.2346   \\
\hline
1  & -0.015  & 0.975  & 200.6905   \\
\hline
1  & -0.03  & 0.975  & 197.7356   \\
\hline
1  & -0.045  & 0.975  & 197.1083   \\
\hline
1  & 0.045 & 0.9825  & 196.8689   \\
\hline
1  & 0.03  & 0.9825  & 193.3416   \\
\hline
1  & 0.015  & 0.9825  & 196.6589   \\
\hline
1  & 0  & 0.9825  & 210.8979   \\
\hline
1  & -0.015  & 0.9825  & 197.0869   \\
\hline
1  & -0.03  & 0.9825  & 195.9922   \\
\hline\hline
1  & -0.045  & 0.9825  & 191.1388   \\
\hline\hline
1  & 0.045  & 1  & 191.6064   \\
\hline
1  & 0.03  & 1  & 198.1114   \\
\hline
1  & 0.015  & 1  & 195.4679   \\
\hline
1  & 0  & 1  & 195.9239   \\
\hline
\end{tabular}
\begin{tabular}{||c c c c||} 
 \hline
$\theta_1$ & $\theta_2$ & $\theta_3$ & CV Error \\ [2ex] 
 \hline\hline
1  & -0.015  & 1  & 201.5761   \\
\hline
1  & -0.03  & 1  & 193.2997   \\
\hline
1  & -0.045  & 1  & 196.1762   \\
\hline
1  & 0.045  & 1.0125  & 194.0597   \\
\hline
1  & 0.03  & 1.0125  & 196.2375   \\
\hline
1  & 0.015  & 1.0125  & 197.0733   \\
\hline
1  & 0  & 1.0125  & 195.814   \\
\hline
1  & -0.015  & 1.0125  & 200.0738   \\
\hline
1  & -0.03  & 1.0125  & 206.5555   \\
\hline
1  & -0.045  & 1.0125  & 195.1094   \\
\hline
1  & 0.045  & 1.025  & 202.3582   \\
\hline
1  & 0.03  & 1.025  & 191.3562   \\
\hline
1  & 0.015  & 1.025  & 204.5542   \\
\hline
1  & 0  & 1.025  & 189.7399   \\
\hline
1  & -0.015  & 1.025  & 203.9044   \\
\hline
1  & -0.03  & 1.025  & 191.532   \\
\hline
1  & -0.045  & 1.025  & 192.4589   \\
\hline
 \hline
\end{tabular}
\end{center}
\renewcommand\thetable{4B}
\caption{CV L2 error (item set: i170, i200, i201, i271)}
\label{table:4}
\end{table}

\begin{table}[H]
\begin{center}
\begin{tabular}{||c c c c||} 
 \hline
$\theta_1$ & $\theta_2$ & $\theta_3$ & CV Error \\ [2ex] 
 \hline\hline
1  & 0.045  & 0.975  & 447.7423   \\
\hline
1  & 0.03  & 0.975  & 447.2207   \\
\hline
1  & 0.015  & 0.975  & 448.0865   \\
\hline
1  & 0  & 0.975  & 448.9228   \\
\hline
1  & -0.015  & 0.975  & 448.5695   \\
\hline
1  & -0.03  & 0.975  & 448.2828   \\
\hline
1  & -0.045  & 0.975  & 450.6857   \\
\hline
1  & 0.045  & 0.9825  & 448.4203   \\
\hline
1  & 0.03  & 0.9825  & 448.0289   \\
\hline
1  & 0.015  & 0.9825  & 449.3513   \\
\hline
1  & 0  & 0.9825  & 451.4599   \\
\hline\hline
1  & -0.015  & 0.9825  & 445.6711   \\
\hline\hline
1  & -0.03  & 0.9825  & 447.1259   \\
\hline
1  & -0.045  & 0.9825  & 447.2945   \\
\hline
1  & 0.045  & 1  & 449.3804   \\
\hline
1  & 0.03  & 1  & 446.9709   \\
\hline
1  & 0.015  & 1  & 449.6299   \\
\hline
1  & 0  & 1  & 451.5521   \\
\hline
\end{tabular}
\begin{tabular}{||c c c c||} 
 \hline
$\theta_1$ & $\theta_2$ & $\theta_3$ & CV Error \\ [2ex] 
 \hline\hline
1  & -0.015  & 1  & 453.9945   \\
\hline
1  & -0.03  & 1  & 449.6805   \\
\hline

1  & -0.045  & 1  & 449.9743   \\
\hline
1  & 0.045  & 1.0125  & 448.3564   \\
\hline
1  & 0.03  & 1.0125  & 451.621   \\
\hline
1  & 0.015  & 1.0125  & 448.5162   \\
\hline
1  & 0  & 1.0125  & 446.6302   \\
\hline
1  & -0.015  & 1.0125  & 448.9254   \\
\hline
1  & -0.03  & 1.0125  & 447.3122   \\
\hline
1  & -0.045  & 1.0125  & 448.591   \\
\hline
1  & 0.045  & 1.025  & 450.3971   \\
\hline
1  & 0.03  & 1.025  & 448.0999   \\
\hline
1  & 0.015  & 1.025  & 449.292   \\
\hline
1  & 0  & 1.025  & 447.8478   \\
\hline
1  & -0.015  & 1.025  & 449.8871   \\
\hline
1  & -0.03  & 1.025  & 454.3573\\
\hline
1  & -0.045  & 1.025  & 449.0327   \\
\hline
1  & 0.45  & 0.9  & 448.2508   \\
\hline
\end{tabular}
\end{center}
\renewcommand\thetable{5A}
\caption{CV L1 error (item set: i102, i160, i172, i202)}
\label{table:5}
\end{table}

\begin{table}[H]
\begin{center}
\begin{tabular}{||c c c c||} 
 \hline
$\theta_1$ & $\theta_2$ & $\theta_3$ & CV Error \\ [2ex] 
 \hline\hline
1  & 0.045  & 0.975  & 217.7359   \\
\hline
1  & 0.03  & 0.975  & 217.7724   \\
\hline
1  & 0.015  & 0.975  & 217.9794   \\
\hline
1  & 0  & 0.975  & 219.3387   \\
\hline
1  & -0.015  & 0.975  & 218.8253   \\
\hline
1  & -0.03  & 0.975  & 218.4537   \\
\hline
1  & -0.045  & 0.975  & 220.2183   \\
\hline
1  & 0.045  & 0.9825  & 218.5969   \\
\hline
1  & 0.03  & 0.9825  & 219.3161   \\
\hline
1  & 0.015  & 0.9825  & 219.0224   \\
\hline
1  & 0  & 0.9825  & 221.2403   \\
\hline\hline
1  & -0.015  & 0.9825  & 216.2027   \\
\hline\hline
1  & -0.03  & 0.9825  & 218.1926   \\
\hline
1  & -0.045  & 0.9825  & 218.7398   \\
\hline
1  & 0.045  & 1  & 218.9043   \\
\hline
1  & 0.03  & 1  & 217.226   \\
\hline
1  & 0.015  & 1  & 220.9737   \\
\hline
1  & 0  & 1  & 220.3873   \\
\hline
\end{tabular}
\begin{tabular}{||c c c c||} 
 \hline
$\theta_1$ & $\theta_2$ & $\theta_3$ & CV Error \\ [2ex] 
 \hline\hline
1  & -0.015  & 1  & 223.7322   \\
\hline
1  & -0.03  & 1  & 220.5439   \\
\hline
1  & -0.045  & 1  & 220.3708   \\
\hline
1  & 0.045  & 1.0125  & 218.5088   \\
\hline
1  & 0.03  & 1.0125  & 221.2157   \\
\hline
1  & 0.015  & 1.0125  & 218.746   \\
\hline
1  & 0  & 1.0125  & 217.2536   \\
\hline
1  & -0.015  & 1.0125  & 218.6351   \\
\hline
1  & -0.03  & 1.0125  & 216.8291   \\
\hline
1  & -0.045  & 1.0125  & 217.5458   \\
\hline
1  & 0.045  & 1.025  & 219.5415   \\
\hline
1  & 0.03  & 1.025  & 218.1658   \\
\hline
1  & 0.015  & 1.025  & 218.3068   \\
\hline
1  & 0  & 1.025  & 218.306   \\
\hline
1  & -0.015  & 1.025  & 220.0956   \\
\hline
1  & -0.03  & 1.025  & 223.4123   \\
\hline
1  & -0.045  & 1.025  & 218.7576   \\
\hline
1  & 0.45  & 0.9  & 218.8855   \\
\hline
\end{tabular}
\end{center}
\renewcommand\thetable{5B}
\caption{CV L2 error (item set: i102, i160, i172, i202)}
\label{table:6}
\end{table}

For the first set of items (which had the least number of NAs among them) (i170, i200, i201, i271) the CV L1 error and CV L2 error is minimized at $\theta=(\theta_1,\theta_2,\theta_3)=(1, -0.03, 1.025)$ \& $(1, -0.045, 0.9825)$ respectively but these points are quite close to $\theta=(1,0,1)$. We conclude from here that if we take ideal case scenario, i.e., $\theta=(1,0,1)$, it works quite well and hence there is not much significant improvement in the model considering measurement error.       

Similarly, for the second set of items (i102, i160, i172, i202) (having second least number of NAs among them) both the CV L1 error and CV L2 error are minimized at $\theta = (\theta_1,\theta_2,\theta_3) = (1, -0.015, 0.9825)$ which is again quite close to $\theta =(1,0,1)$. Again for this set, ignoring the measurement error won't have any significant effect on the conclusions.

Thus, overall, measurement error does not play any significant role in distorting the conclusions of our model and methodology.

\section{Analysis of Inequality: Lorenz Curve and Gini Index}

In the analysis above, we studied whether the alternate dataset, as explained in section 3.2, is similar to the original dataset in terms of budget share distribution. In this section we take the analysis one step further to check whether CE inequality metrics are influenced by our alternative methodology or not. We will examine this using the Lorenz Curve and the Gini Index.

We first fix an item $i.$ Corresponding to item $i$ we have information about it's price $p_{i_j}$ in the $j^{th}$ household. We also know exact value of amount of the item $i$ bought in the $j^{th}$ household, $q_{i_j}=\frac{w_{i_j}Q_j}{p_{i_j}}$ where $w_{i_j}$ and $Q_j$ are respectively the share of item $i$ and total expenditure in the $j^{th}$ household. Now we draw the Lorenz curve for expenditure for item i using this original dataset.
\\Next we do the same but for $\Tilde{p_{i_j}}$, the constant price for $j^{th}$ household corresponding to the FSU it belongs to. $q_{i_j}$ remains same here as well. We draw the Lorenz curve for the expenditure using the alternate dataset and then we compare it with previous curve. This is done for all the items that we had worked with (items $170, 200, 201, 271, 102, 160, 172, 202$). We also compute the Gini index for these two alternatives for comparison.

The Lorenz curves for item no. 102, 170, 172, 200 are in figures 1 - 8. The right panel is for the original dataset and the left ones for alternate. As we can see, the Lorenz curves for the alternate and the original data are very close to each other, implying that our simplification do not result in any distortion.

\begin{figure}[H]
\centering
\begin{minipage}{.5\textwidth}
  \centering
 \includegraphics[width=\linewidth]{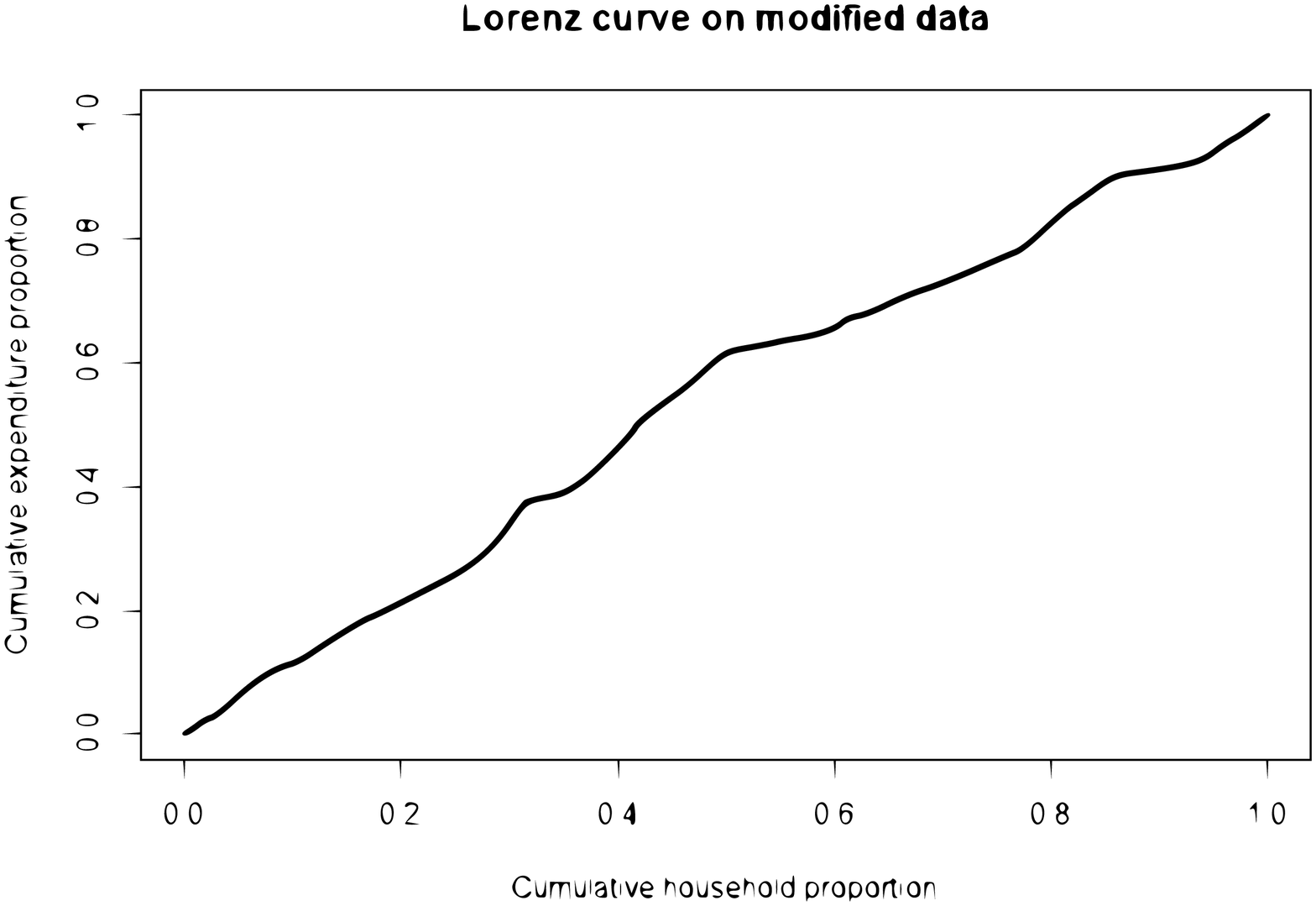}
  \caption{i102 alternate data}
  \label{fig:test1}
\end{minipage}%
\begin{minipage}{.5\textwidth}
  \centering
  \includegraphics[width=\linewidth]{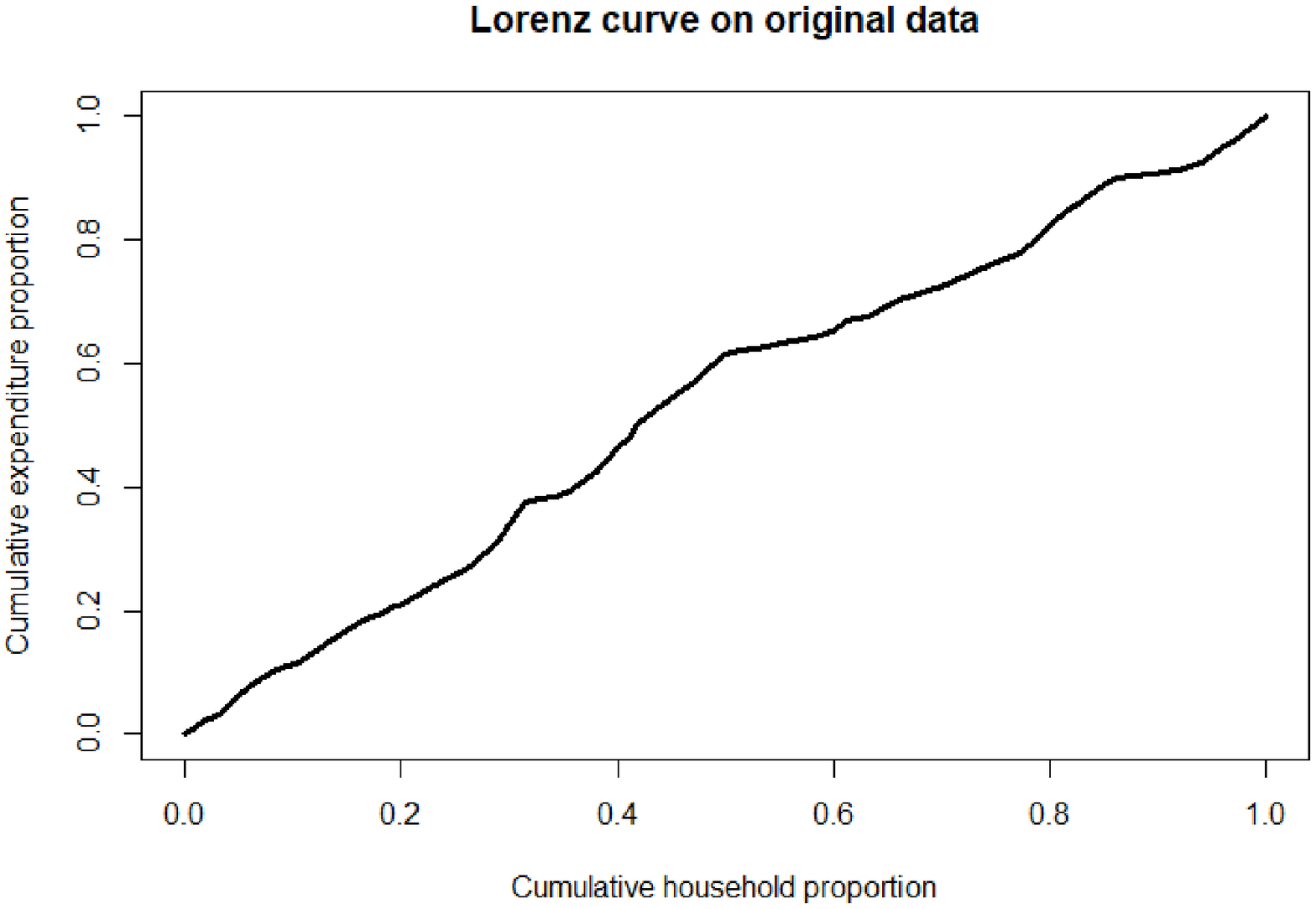}
  \caption{i102 original data}
  \label{fig:test2}
\end{minipage}
\end{figure}

\begin{figure}[H]
\centering
\begin{minipage}{.5\textwidth}
  \centering
  \includegraphics[width=\linewidth]{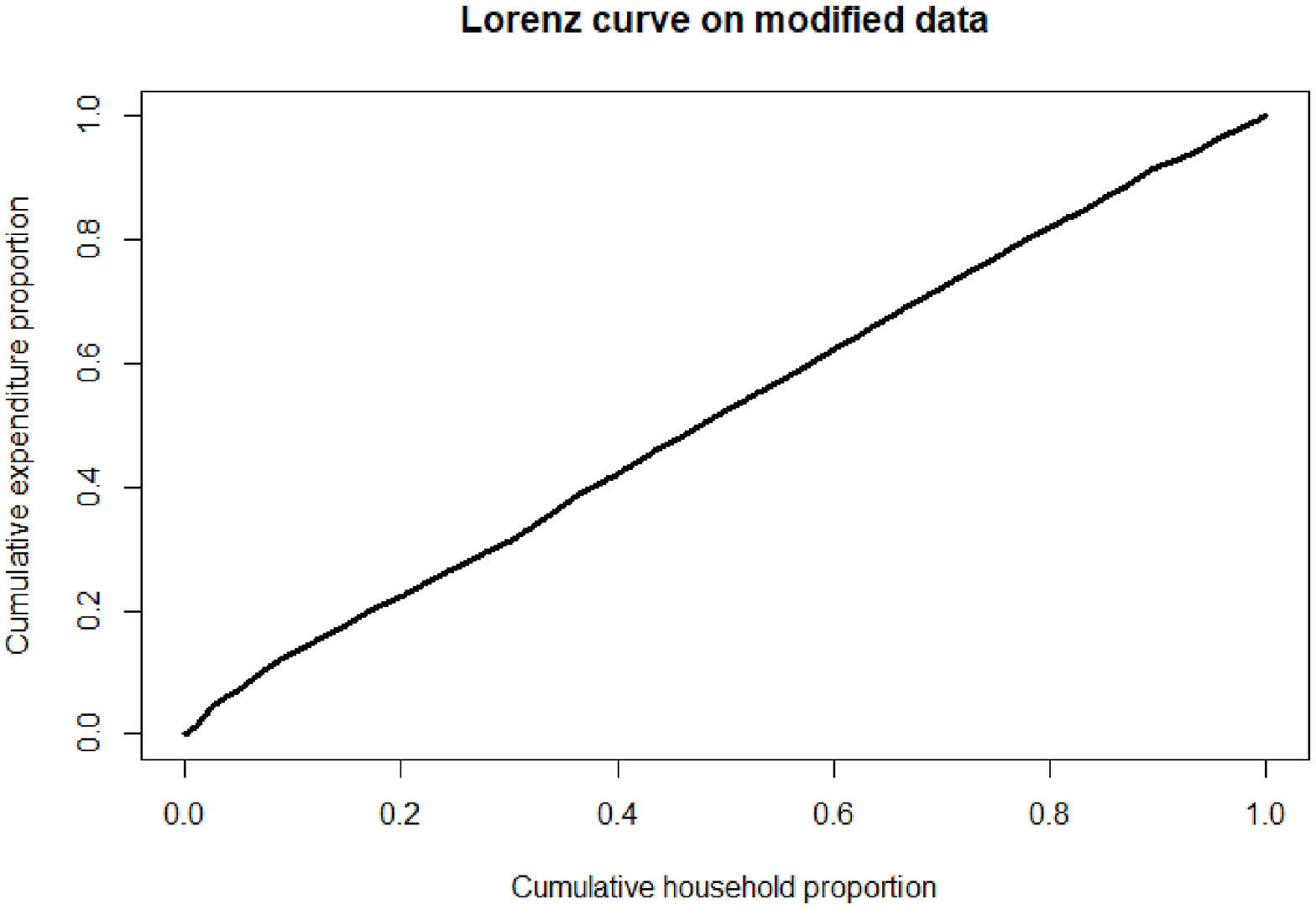}
  \caption{i170 alternate data}
  \label{fig:test1}
\end{minipage}%
\begin{minipage}{.5\textwidth}
  \centering
  \includegraphics[width=\linewidth]{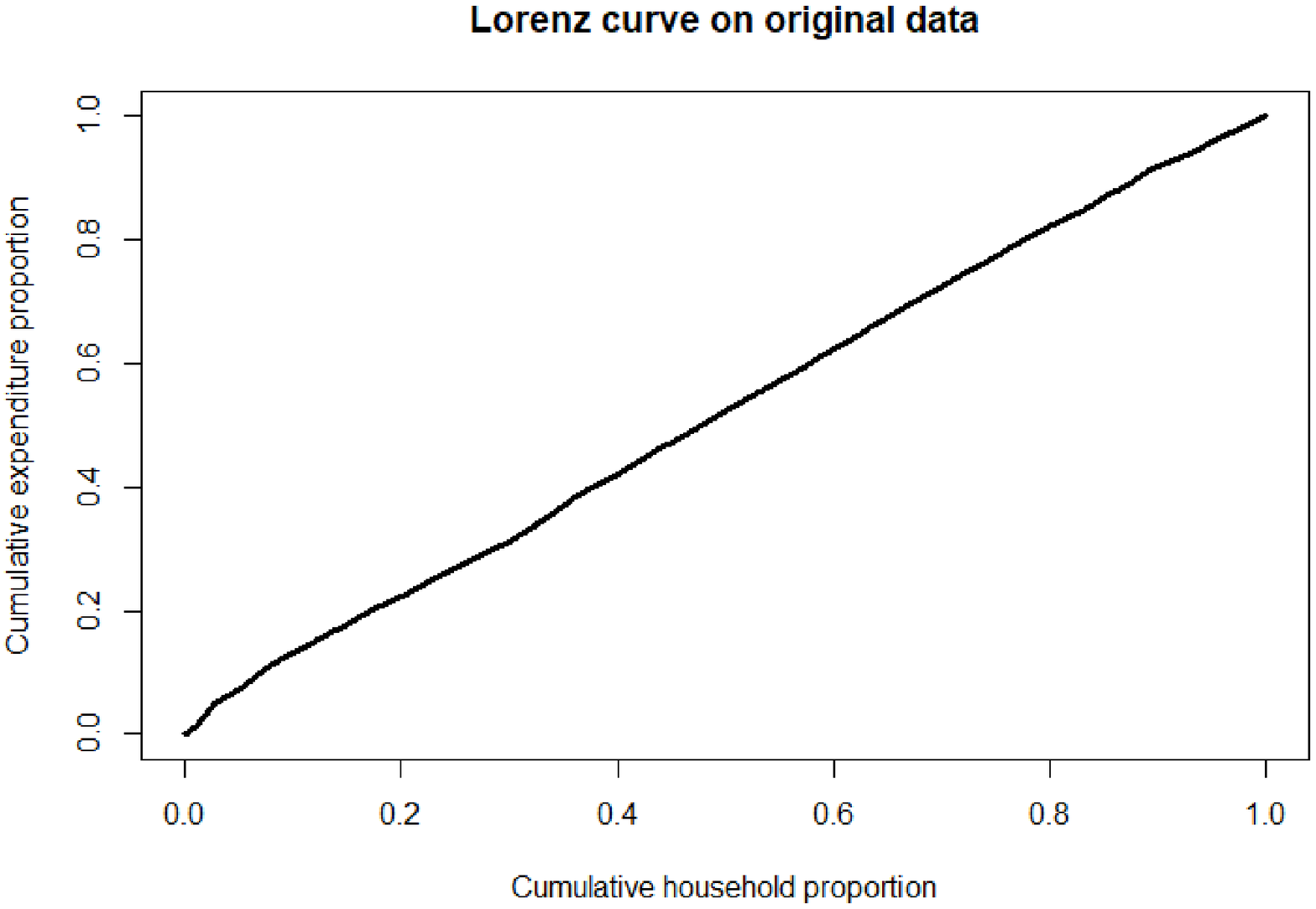}
  \caption{i170 original data}
  \label{fig:test2}
\end{minipage}
\end{figure}

\begin{figure}[H]
\centering
\begin{minipage}{.5\textwidth}
  \centering
  \includegraphics[width=\linewidth]{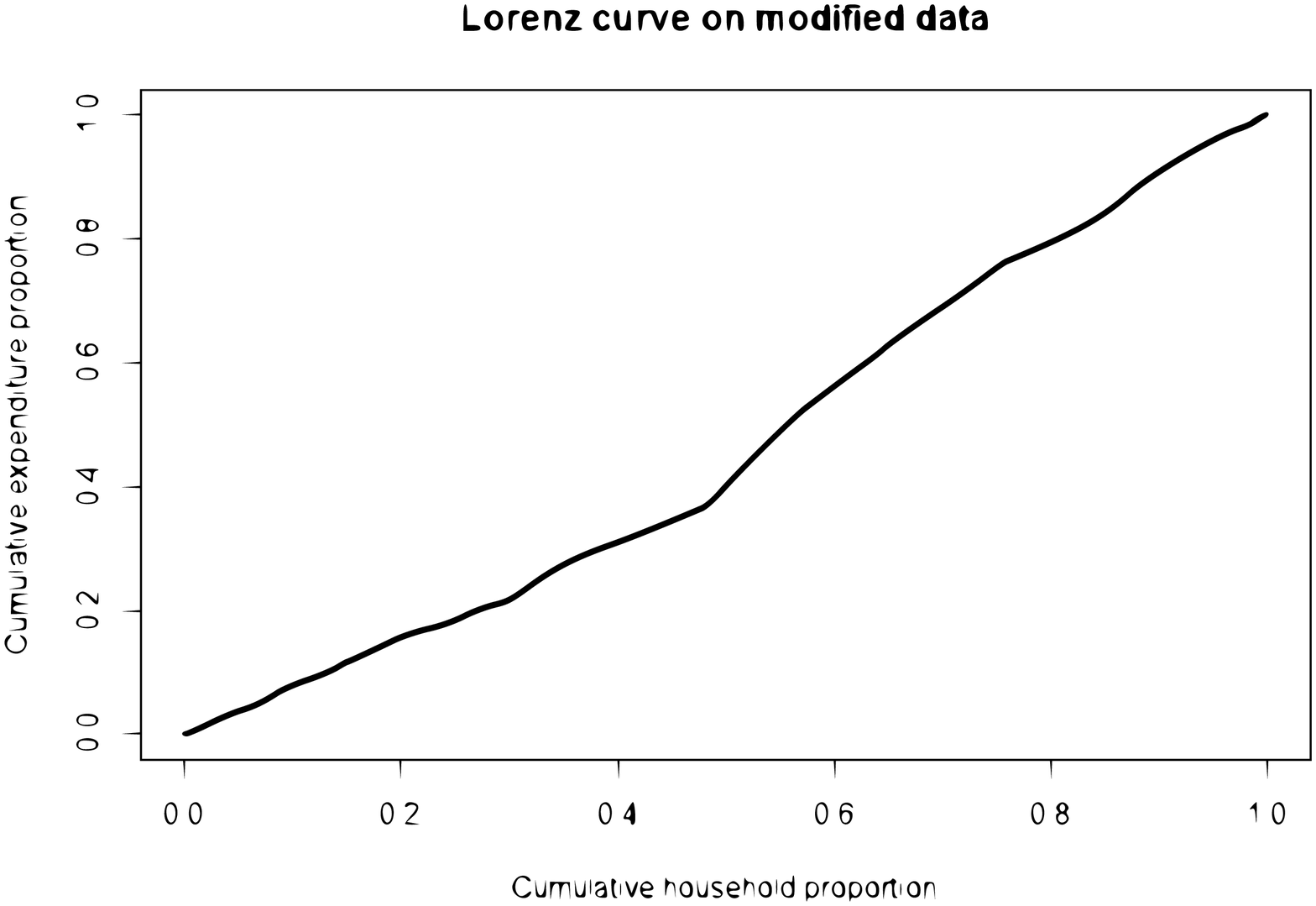}
  \caption{i172 alternate data}
  \label{fig:test1}
\end{minipage}%
\begin{minipage}{.5\textwidth}
  \centering
  \includegraphics[width=\linewidth]{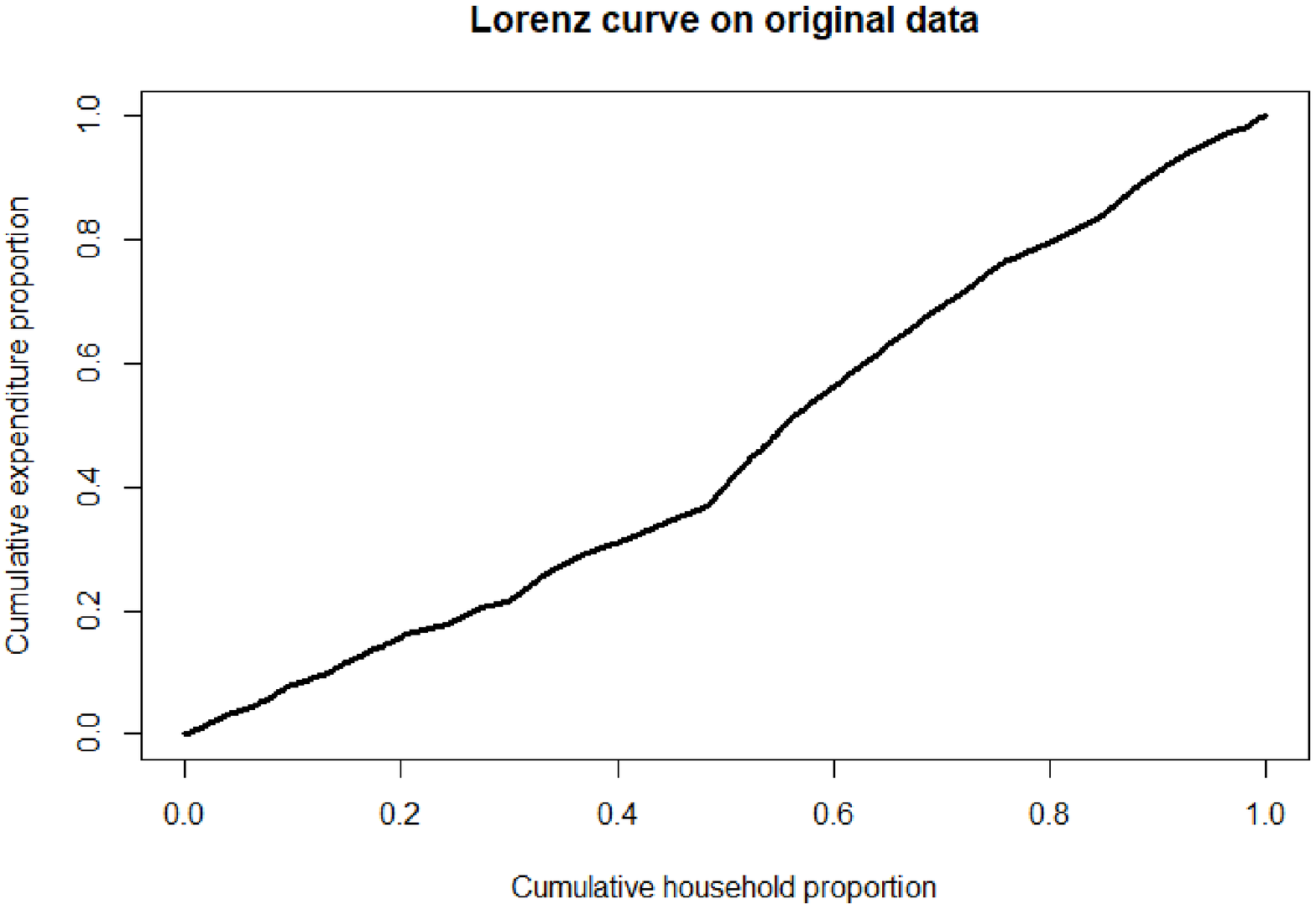}
  \caption{i172 original data}
  \label{fig:test2}
\end{minipage}
\end{figure}

\begin{figure}[H]
\centering
\begin{minipage}{.5\textwidth}
  \centering
  \includegraphics[width=\linewidth]{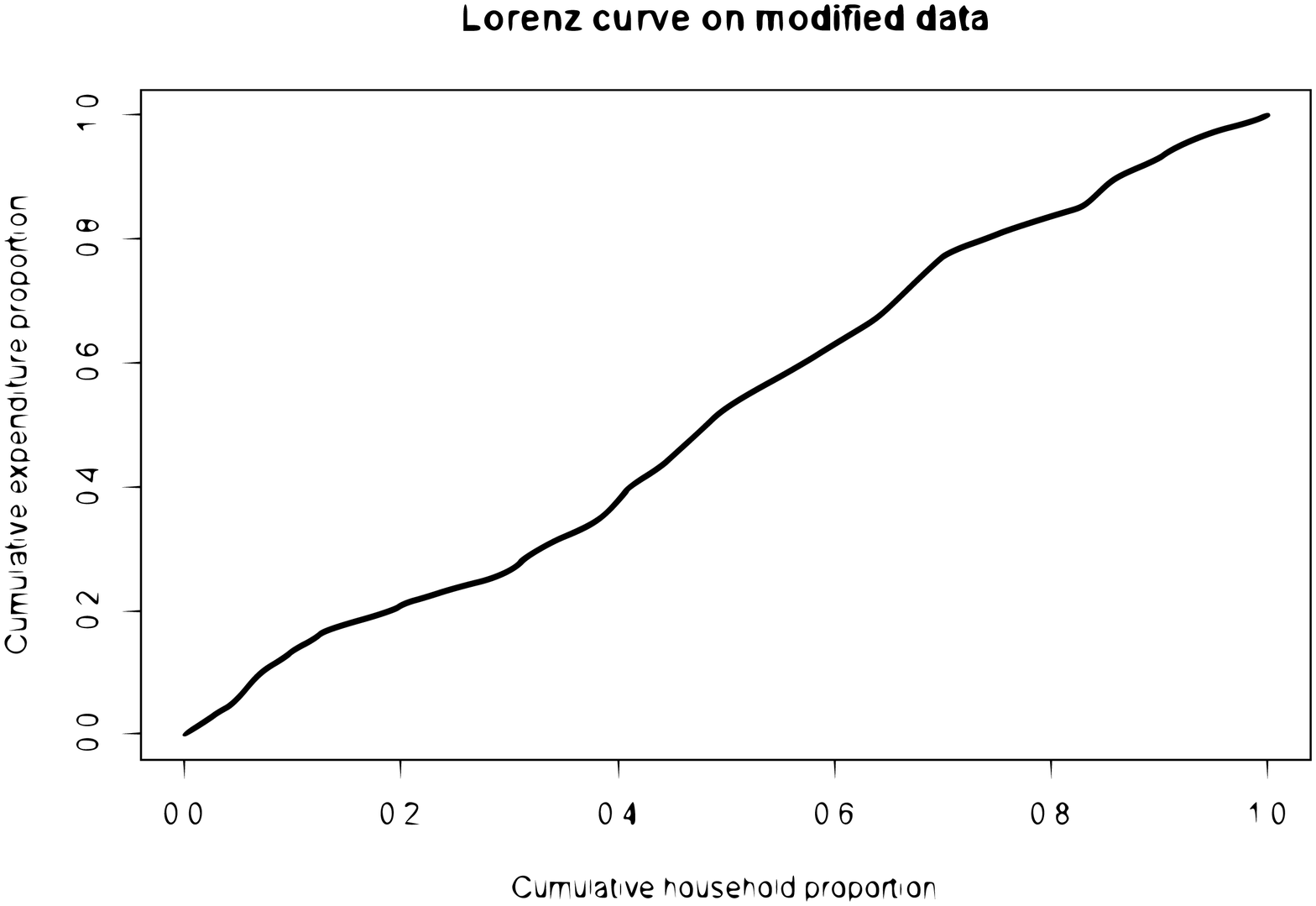}
  \caption{i200 alternate data}
  \label{fig:test1}
\end{minipage}%
\begin{minipage}{.5\textwidth}
  \centering
  \includegraphics[width=\linewidth]{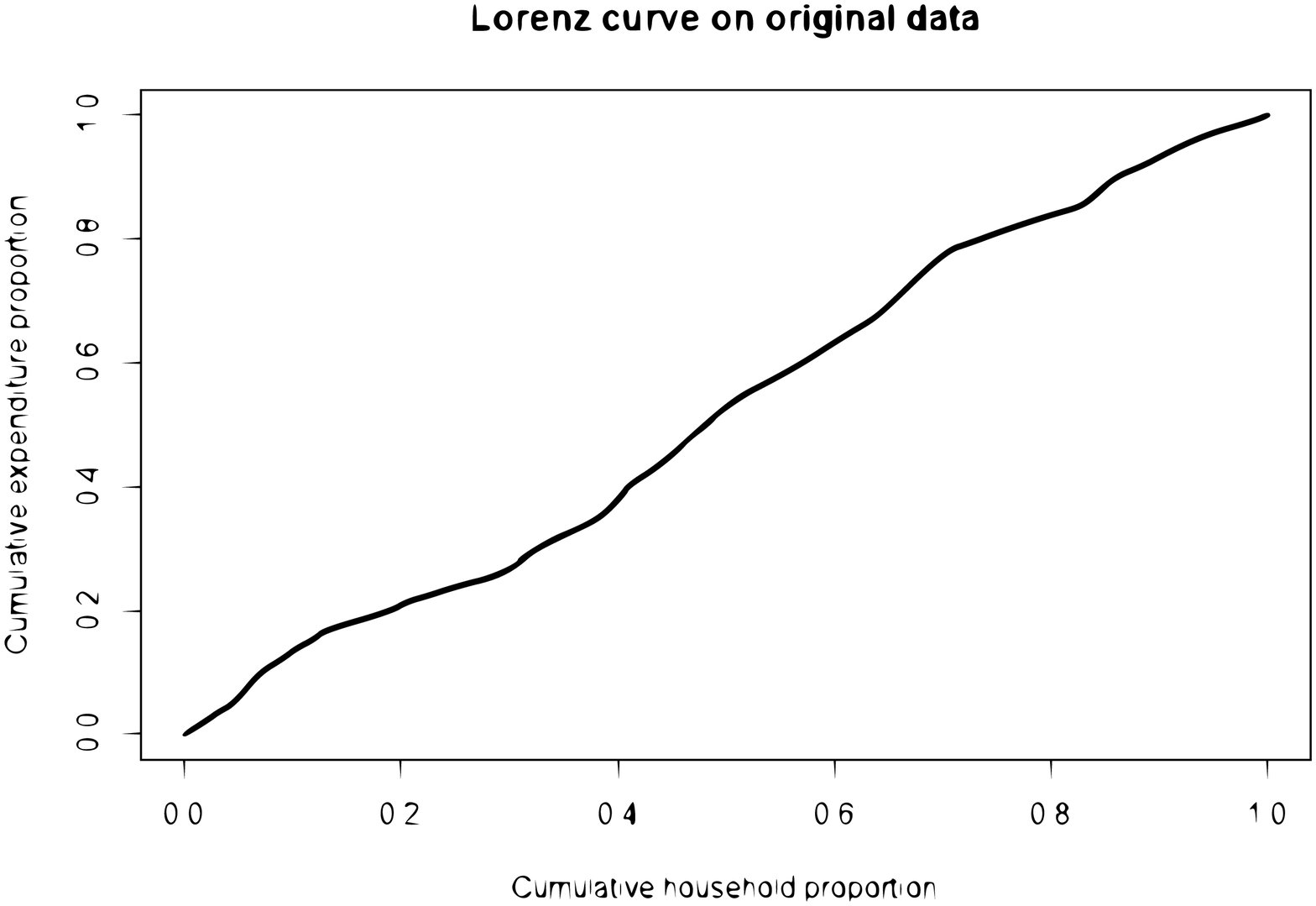}
  \caption{i200 original data}
  \label{fig:test2}
\end{minipage}
\end{figure}

The Gini index for the 2 datasets is as follows (Data 1 is original, Data 2 is alternate):
\begin{center}
\begin{tabular}{||c c c|c|} 
 \hline
 item no. & Data 1 & Data 2 & $|\Phi(T_1)-\Phi(T_2)|$\\ [0.5ex] 
 \hline\hline
 102 & 0.447 & 0.452 & 0.05\\ 
 \hline
 170 & 0.290 & 0.298 & 0\\
 \hline
 172 & 0.370 & 0.370 & 0.001\\
 \hline
 200 & 0.407 & 0.418 & 0\\
  \hline
 201 & 0.358 & 0.366 & 0\\
  \hline
 202 & 0.383 & 0.394 & 0\\
  \hline
 271 & 0.390 & 0.408 & 0\\
 \hline
 160 & 0.411 & 0.409 & 0.003 \\ [1ex] 
 \hline
\end{tabular}

~\\
Table 6: Gini index comparison
\end{center}

We make a comparison about the $significance$ of gini index difference among the 8 items. As discussed in  \cite{martinez2009exact}, if $U=\{u_1,u_2,...,u_n\}$ is a random sample from a $U(0,1)$ distribution, then 
\[ \sqrt{n} \frac{\hat{G_n}-1/3}{\sqrt{8/135}} \xrightarrow[]{d} N(0,1)\]
where $\hat{G_n}$ is sample Gini index.
\newline Since the data of expenditure for a item in a particular household does not necessarily follow uniform distribution, we can not apply the above result directly. We rather calculate the empirical distribution $\hat{F_k}$ of the expenditure of each household for item $i$ ($k$ stands for the data used). Next we define $u_j=\hat{F_k}(x_j)$ where $x_j$ is the expenditure of $j^{th}$ household on item $i$ ($u_j \sim U(0,1)$). Now we calculate the test statistic $T_k=\sqrt{n} \frac{\hat{G}_{nk}-1/3}{\sqrt{8/135}}$ for both the data sets and check the value of $|\Phi(T_1)-\Phi(T_2)|$, where $\Phi$ is the CDF of the standard Normal distribution. If this difference is more that $0.05$, we conclude that the Gini index due to item $i$ are not statistically same for both the datas.
For all our items the difference turns out to be insignificant, again proving that Data 1 and Data 2 are statistically similar to each other w.r.t. inequality in the expenditure distribution.

\section{Effect on Demand Elasticity}

We conclude our analysis with a comparison of demand and price elasticities (both Marshallian and Hicksian) under the AIDS model between the original and the alternate dataset.

\begin{table}[H]
\begin{center}
\begin{tabular}{||c c c|c|} 
 \hline
  Item number & modified dataset & original dataset & absolute difference ($10^{-3}$) \\
 \hline\hline
 i102 & 1.3095478& 1.3036325 & 5.9153\\ 
 \hline
 i170 & 0.441356 & 0.4351241 & 6.2319\\ 
 \hline
 i172 & 0.3912798 & 0.4053625 & \textbf{14.0827}\\
 \hline
 i200 & 0.3165078 & 0.3040078 & 12.5\\
 \hline
 i202 & 0.2207938 & 0.2085229 & 12.2709\\
 \hline
 i271 & 0.2225349 & 0.2231390 & 0.6041\\
 \hline
 i160 & 0.3229050 & 0.3173401 & 5.5649\\
 \hline
 i220 & 1.0817749 & 1.0806154 & 1.1595\\ [1ex] 
 \hline
\end{tabular}
\end{center}
\renewcommand\thetable{7}
\caption{Demand elasticities of 8 Items}
\label{table:demand elasticities}
\end{table}

\begin{table}[!htb]
\begin{minipage}{.45\linewidth}
    \centering

    \label{tab:first_table}

    \medskip
\begin{adjustbox}{scale=0.9}
\begin{tabular}{| c c c  c c |} 
\toprule
\makecell{ Marshallian}  &  p\_102 & p\_170 & p\_172 & p\_200 \\   
\midrule
q\_i102 & -1.761 & 0.013 & 0.391 & 0.054\\
q\_i170 & 0.898 & -0.553 & -0.657 & -0.117   \\
q\_i172 & 1.593 & -0.070 & -1.825 & -0.103   \\
q\_i200 &  1.481 & -0.071 & -0.585 & \textcolor{red}{-1.129} \\
\bottomrule
\toprule
\makecell{}  &  p\_202 & p\_271 & p\_160 & p\_220 \\   
\midrule
q\_i202 & \textcolor{red}{-0.523} & -0.141 & -0.168 & 0.629\\
q\_i271 & -0.132 & \textcolor{red}{-0.581} & -0.098 & 0.588   \\
q\_i160 & -0.106 & -0.067 & \textcolor{red}{-0.744} & \textcolor{red}{0.600}   \\
q\_i220 &  -0.004 & -0.006 & -0.004 & -1.066 \\
\bottomrule

\end{tabular}
\end{adjustbox}
\end{minipage}\hfill
\begin{minipage}{.45\linewidth}
    \centering

    \label{tab:second_table}

    \medskip
\begin{adjustbox}{scale=0.9}
\begin{tabular}{| c c c  c c |} 

\toprule
\makecell{ Hicksian}  &  p\_102 & p\_170 & p\_172 & p\_200 \\   
\midrule
q\_i102 & -0.893 & 0.050 & 0.732 & 0.112\\
q\_i170 & 1.182 & -0.541 & -0.543 & -0.100   \\
q\_i172 & 1.863 & -0.058 & -1.720 & -0.085   \\
q\_i200 &  1.683 & -0.062 & -0.506 & \textcolor{blue}{-1.115} \\
\bottomrule
\toprule
\makecell{}  &  p\_202 & p\_271 & p\_160 & p\_220 \\   
\midrule
q\_i202 & \textcolor{blue}{-0.522} & -0.135 & -0.159 & 0.817\\
q\_i271 & -0.126 & \textcolor{blue}{-0.575} & -0.089 & 0.790   \\
q\_i160 & \textcolor{blue}{-0.010} & -0.058 & \textcolor{blue}{-0.730} & \textcolor{blue}{0.886}   \\
q\_i220 &  0.024 & 0.025 & -0.043 & -0.095 \\
\bottomrule

\end{tabular}
\end{adjustbox}
\end{minipage}
\renewcommand\thetable{8A}
\caption{Original Dataset}
\end{table}

\begin{table}[!htb]
\begin{minipage}{.45\linewidth}
    \centering

    \label{tab:first_table}

    \medskip
\begin{adjustbox}{scale=0.9}
\begin{tabular}{| c c c  c c |} 
\toprule
\makecell{Marshallian }  &  p\_102 & p\_170 & p\_172 & p\_200 \\   
\midrule
q\_i102 & -1.758 & 0.014 & 0.382 & 0.052\\
q\_i170 & 0.906 & -0.467 & -0.749 & -0.132   \\
q\_i172 & 1.580 & -0.079 & -1.777 & -0.115   \\
q\_i200 &  1.435 & -0.079 & -0.654 & \textcolor{red}{-1.018} \\
\bottomrule
\toprule
\makecell{}  &  p\_202 & p\_271 & p\_160 & p\_220 \\   
\midrule
q\_i202 & \textcolor{red}{-0.395} & -0.180 & -0.209 & 0.563\\
q\_i271 & -0.165 & \textcolor{red}{-0.460} & -0.129 & 0.531   \\
q\_i160 & -0.129 & -0.088 & \textcolor{red}{-0.522} & \textcolor{red}{0.416}   \\
q\_i220 &  -0.006 & -0.007 & -0.013 & -1.054 \\
\bottomrule

\end{tabular}
\end{adjustbox}
\end{minipage}\hfill
\begin{minipage}{.45\linewidth}
    \centering

    \label{tab:second_table}

    \medskip
\begin{adjustbox}{scale=0.9}
\begin{tabular}{| c c c  c c |} 
\toprule
\makecell{Hicksian }  &  p\_102 & p\_170 & p\_172 & p\_200 \\   
\midrule
q\_i102 & -0.887 & 0.051 & 0.725 & 0.111\\
q\_i170 & 1.200 & -0.455 & -0.633 & -0.111   \\
q\_i172 & 1.840 & -0.068 & -1.674 & -0.098   \\
q\_i200 & 1.646 & -0.070 & -0.571 & \textcolor{blue}{-1.003} \\
\bottomrule
\toprule
\makecell{}  &  p\_202 & p\_271 & p\_160 & p\_220 \\   
\midrule
q\_i202 & \textcolor{blue}{-0.389} & -0.173 & -0.200 & 0.761\\
q\_i271 & -0.159 & \textcolor{blue}{-0.454} & -0.119 & 0.731   \\
q\_i160 & \textcolor{blue}{-0.120} & -0.079 & \textcolor{blue}{-0.508} & \textcolor{blue}{0.710}   \\
q\_i220 &  0.023 & 0.024 & -0.035 & -0.081 \\
\bottomrule

\end{tabular}
\end{adjustbox}
\end{minipage}
\renewcommand\thetable{8B}
\caption{Modified dataset}
\end{table}

As we can clearly see, the expenditure elasticities are very similar for both the data sets ($0.014$ being the maximum difference). Most of the price elasticities are also close to each other and all are of the same sign. \textcolor{blue}{Blue} marked ones are the Marshallian price elasticities which have an absolute difference larger than $0.1$ between the two datasets and \textcolor{red}{Red} marked ones are the Hicksian price elasticities which have an absolute difference larger than $0.1$ between the two datasets with the maximum being 0.222. But the average difference is of a much lower order (around $0.04$).
So we can conclude that our modified alternate dataset is a reasonable simplification, without any significant loss of information,  of the original dataset.

\section{Conclusion}
\label{sec:conclusions}
Consumer Expenditure surveys are of paramount importance for macro planning for a country. Huge amount of money is spent on this every year. It would be an important contribution if a modification to existing sampling design leads to a reduction in cost with no appreciable loss of precision.
This paper showed that utilising the model structure corresponding to uniform prices within FSUs does not lead to a significant loss in information for quantifying the consumer expenditure distribution as well as the associated shares as compared to the previous methodology of complete information, which in turn motivates us towards using the uniform prices setup during surveys to help in cost reduction and optimisation. We used a modified version of the AIDS (Almost Ideal Demand System) structure for quantifying the relationships between quantities of goods, total expenditure and prices under the influence of state effects and measurement errors. 
We have employed a Cross-Validation based technique to incorporate the optimum measurement error model for prices and total expenditure during the survey, and thus provide a generalised model structure. Due to data issues we have only been able to do the analysis for a few items. But the results were uniformly encouraging, showing no loss of information - except for one item - and no significant measurement error issues. Thus, this work paves a way towards introducing a robust and cost efficient survey methodology for Consumer Expenditure. 

Our work may be extended in several directions in future. Using other general demand systems may indicate robustness of results w.r.t. model specification. Applications in the direction of measuring Poverty or Human Development may also be important in terms of policy implications. We hope that such generalisations and applications will enrich our knowledge further in this direction in the near future. \\

\noindent {\bf Funding source:} 
This research did not receive any specific grant from funding agencies in the public,  commercial,  or not-for-profit sectors. \\      

\noindent {\bf Declarations of interest:} 
The authors declare no conflict of interest.

\bibliography{Bibliography-MM-MC}
\end{document}